\documentclass[aps,prb,twocolumn,amsmath,showpacs,amssymb,superscriptaddress]{revtex4}

\usepackage{amsmath}
\usepackage{graphicx}
\usepackage{amssymb}
\usepackage{dsfont}
\usepackage{color}

\newcommand{\pdag}{{\phantom{\dagger}}}

\begin{document}

\title{Fractional topological phases in three-dimensional coupled-wire systems}

\author{Tobias Meng}
\affiliation{Institut f\"ur Theoretische Physik, Technische Universit\"at Dresden, 01062 Dresden, Germany}

\begin{abstract}
It is shown that three-dimensional systems of coupled quantum wires support fractional topological phases composed of closed loops and open planes of two-dimensional fractional quantum Hall subsystems. These phases have topologically protected edge states, and are separated by exotic quantum phase transitions corresponding to a rearrangement of fractional quantum Hall edge modes. Some support for the existence of an extended exotic critical phase separating the bulk gapped fractional topological phases is given. Without electron-electron interactions, similar but unfractionalized bulk gapped phases based on coupled integer quantum Hall states exist. They are separated by an extended critical Weyl semimetal phase.
\end{abstract}

\maketitle

\section{Introduction}
Since the experimental discovery and theoretical explanation of the fractional quantum Hall effect,\cite{tsui_fqhe,laughlin_83} fractionalization in interacting topological systems has been an important theme in condensed matter physics. While the understanding of fractionalized topological phases has impressively developed in two dimensions (2D), much less is known in three dimensions (3D). Slave-particle approaches have for instance allowed one to make progress for topological Mott insulators,\cite{pesin_10, wk_10,scheurer_15} and 3D fractional topological insulators.\cite{maciejko_10,swingle_11,maciejko_12,lee_12} Some exactly solvable models have been reported,\cite{castelnovo_08,levin_11} and the Kitaev honeycomb model\cite{kitaev_06} has been generalized to 3D lattices.\cite{si_08,mandal_09,ryu_09,wu_09,hermanns_14,hermanns_15} Further examples for fractionalized 3D phases include spin ice,\cite{castelnovo_07,morris_09,fennel_09,bramwell_09} and stacks of 2D fractional quantum Hall layers.\cite{balents_96} In the latter, inter layer couplings can stabilize many-layer versions of Halperin bilayer states,\cite{halperin_83,qiu_89,jaud_00} or exotic phases with fractionally charged fermionic 3D quasiparticles.\cite{levin_09} Other coupled-layer constructions \cite{wang_13,jian_14} can also exhibit string-like excitations, which in general allow for non-trivial 3D braiding.\cite{wang_14,jian_14,wang_15,Jiang_14,moradi_15}

Instead of coupling extended layers, this work engineers 3D fractional topological states by connecting 2D building blocks of finite width along different directions. While coupled-layer physics can be recovered by dominantly coupling the building blocks along planes, the blocks may also connect along other geometries, such as closed loops. This gives rise to additional topological phases and phase transitions. For concreteness, the remainder studies narrow integer and fractional quantum Hall strips as building blocks. As a further difference to previous studies, the individual building blocks are constructed from coupled quantum wires containing spin-polarized electrons. Time-reversal symmetry is thus broken from the outset. Since each wire can be treated as a Luttinger liquid,\cite{sondhi_00} this approach is especially powerful for the analysis of interacting phases. Starting with the pioneering work of Kane \textit{et al.},\cite{kane_02,teo_14} it has been shown that coupled-wire constructions allow for an analytically tractable description of integer and fractional, Abelian and non-Abelian topological 2D states.\cite{lu-12prb125119,vaezi-14prl236804,seroussi-14prb104523,klinovaja-14epjb87,meng-14epjb203,sagi-14prb201102,mong-14prx011036,vaezi14prx031009, klino_14,klino_15,santos_15,meng_15,goroh_15} Coupled-wire constructions have also led to qualitatively new results, including a classification of interacting topological phases,\cite{neupert-14prb205101} and the prediction of spontaneously time-reversal-symmetry-broken states towards which 2D fractional topological insulators can be unstable.\cite{meng-14prb235425}

The present work extends this list by adapting coupled-wire constructions as a tool for the analysis of novel interacting topological physics in 3D. The potential of this approach is exemplified by constructing a system that hosts several fractional topological phases. The coupled-wire construction provides simple illustrations of these phases in terms of closed loops (or cylinders) and open planes of integer and fractional quantum Hall subsystems. It also allows one to identify a regime in which an extended exotic critical phase may exist. At special Luther-Emery-type points, the low-energy theory of the system can furthermore be solved by an exact mapping to non-interacting fermions of fractional charge.

The plan of the paper is as follows. Section \ref{sec:blocks} discusses the individual building blocks that are used to construct the 3D system. Section \ref{sec:non_int_array} is devoted to the analysis of the non-interacting array, whose phase diagram is detailed in Sec.~\ref{sec:non_int_phases}. In Sec.~\ref{sec:interactions_ll}, I turn to the description of the interacting 3D system in terms of coupled Luttinger liquids. The interacting phase diagram is discussed in Sec.~\ref{sec:interact_phases}. The results are finally summarized in Sec.~\ref{sec:summary}.

\section{Integer and fractional quantum Hall building blocks}\label{sec:blocks}
\begin{figure}
  \centering
  \includegraphics[width=\columnwidth]{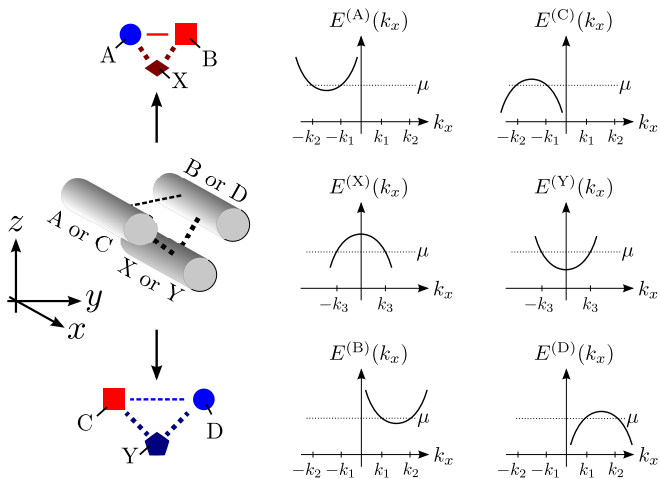}
  \caption{The building blocks A-X-B and C-Y-D, and the dispersions $E^{(\cdot)}(k_x)$ of the different wires. On the left, strong (weak) tunneling couplings are indicated by thick (thin) lines.}
  \label{fig:blocks}
\end{figure}

The Abelian, but in general fractionalized, phases studied in the remainder are constructed from two fundamental building blocks A-X-B and C-Y-D shown in Fig.~\ref{fig:blocks}. The dispersions of the electron-type wires A and B, and the hole-type wires C and D, are asymmetrical with respect to zero momentum $k_x$ along the wires. For a given chemical potential $\mu$, the Fermi points reside at momentum $k_x=-k_{1,2}$ in wires A and C, and $k_x=+k_{1,2}$ in wires B and D. This can be realized in spinful wires with spin-orbit coupling, which are polarized by a magnetic field parallel to the spin-orbit direction. The central wire X is of hole-type, while Y is of electron-type. They have Fermi points at $k_x=\pm k_3$. In each building block, the close-by inner and outer wires are tunnel-coupled by a strong hopping term. A direct tunneling, albeit of reduced strength, also exists between the more distant outer wires.

Reference \onlinecite{sondhi_00} showed that the building blocks form narrow integer quantum Hall strips of opposite chirality if $k_1=0$ and $k_2=k_3$. This can be motivated by noting that the dominant tunnelings between the inner and outer wires induce large gaps for the counterpropagating modes at $k_2=k_3$, and $-k_2=-k_3$, such that the central wire X (Y) forms the bulk of a three wire wide integer quantum Hall strip. A right-moving edge mode of momentum $k_x\approx k_1=0$ lives in wire A (D), while a left-moving edge mode is located in wire B (C). In an isolated building block, these edge modes acquire a small gap due to both their overlap across the central wire, and the direct tunneling between the outer wires.

With electron-electron interactions, and if the filling is reduced, the building blocks can enter fractional quantum Hall states.\cite{kane_02} A Laughlin state at an effective filling factor $\nu=1/(2m+1)$ with positive integer $m$ can for instance arise for $k_1=m k_3$ and $k_2 = (m+1) k_3$. While single-particle tunneling between counterpropagating modes is then forbidden by momentum conservation, correlated tunnelings can drive the building blocks into fractional quantum Hall states. The central wire X (Y) then constitutes the gapped bulk of a now fractional quantum Hall strip, while the outer wires again host chiral edge modes. A more detailed discussion of the physics within a given building block in terms of coupled Luttinger liquids is given in Sec.~\ref{subsec:blocks} below.

\section{3D system of non-interacting wires}\label{sec:non_int_array}
\begin{figure}
  \centering
  \includegraphics[width=0.7\columnwidth]{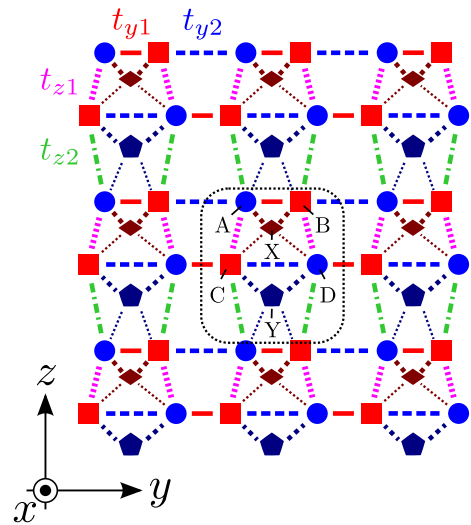}
  \caption{Section of the periodic 3D system (the dotted box shows the unit cell). Thick dotted diagonal lines depict the dominant hoppings in each A-X-B and C-Y-D building block, thin dotted diagonal lines indicate subleading hoppings between the X and C, D (Y and A, B) wires. The edge mode couplings are $t_{y1}$ along solid lines, $t_{y2}$ along dashed lines, $t_{z1}$ along dotted vertical lines, and $t_{z2}$ along dash-dotted lines.}
  \label{fig:system}
\end{figure}

To realize 3D phases, I consider the periodic array shown in Fig.~\ref{fig:system}. I require the tunnelings between the inner and outer wires in each building block to be the largest energy scale after the bandwidth of the wires, and the chemical potential. Additional tunnelings between wires X and C, and X and D within a unit cell, as well as between wires Y and A, and Y and B in neighboring unit cells, which compete with the dominant intra building block tunnelings, can then be neglected, and the low-energy physics of the array is fully described by the integer or fractional quantum Hall edge modes in wires A, B, C, and D, as well as the interactions and tunnelings between them. 

Focussing first on the integer quantum Hall case $k_1=0$, $k_2=k_3$, the low-energy dispersions of the edge modes are well approximated by $\pm v_F k_x$, where $v_F$ denotes the Fermi velocity. Along $y$, I assume neighboring edge modes to be coupled by small alternating hoppings $t_{y1}$ and $t_{y2}$, whose strengths can be controlled by the inter wire distances. I take these hoppings to be shifted by one edge mode in the next $(x,y)$ layer. Along $z$, I connect neighboring edge modes by small tunnelings $t_{z1}$ within the unit cell, and $t_{z2}$ between two adjacent unit cells. All tunnel couplings are chosen to be positive.

Labelling the unit cells by an index $p$ in $y$ direction and $q$ in  $z$ direction, the array is described by the low-energy Hamiltonian
\begin{align}
H&=\sum_{k_x}\sum_{p,q,p'q'}\Psi_{k_xpq}^\dagger\,\begin{pmatrix}\mathcal{H}_{11}\delta_{q,q'}&\mathcal{H}_{12}\delta_{p,p'}\\\mathcal{H}_{21}\delta_{p,p'}&\mathcal{H}_{22}\delta_{q,q'}\end{pmatrix}\,\Psi_{k_xp'q'}^\pdag~,\label{eq:ham}\\
\mathcal{H}_{11}&=\begin{pmatrix}v_F k_x\delta_{p,p'}&t_{y1}\delta_{p,p'}+t_{y2}\delta_{p,p'+1}\\t_{y1}\delta_{p,p'}+t_{y2}\delta_{p,p'-1}&-v_F k_x\delta_{p,p'}\end{pmatrix},\\
\mathcal{H}_{12}&=(t_{z1}\delta_{q,q'}+t_{z2}\delta_{q,q'-1})\,\mathds{1}_{2\times2}~,\\
\mathcal{H}_{21}&=(t_{z1}\delta_{q,q'}+t_{z2}\delta_{q,q'+1})\,\mathds{1}_{2\times2}~,\\
\mathcal{H}_{22}&=\begin{pmatrix}-v_F k_x\delta_{p,p'}&t_{y2}\delta_{p,p'}+t_{y1}\delta_{p,p'+1}\\t_{y2}\delta_{p,p'}+t_{y1}\delta_{p,p'-1}&v_F k_x\delta_{p,p'}\end{pmatrix},
\end{align}
where $\Psi_{k_xpq}=(c_{k_xpq}^{(\rm{A})},c_{k_xpq}^{(\rm{B})},c_{k_xpq}^{(\rm{C})},c_{k_xpq}^{(\rm{D})})^T$ is the vector of annihilation operators for edge mode electrons with momentum $k_x$ in wire A, B, C, and D of unit cell $(p,q)$. This Hamiltonian is essentially identical to the low-energy description of stacked topological insulators analyzed by Burkov and Balents.\cite{burkov_balents_layers_11} Following their calculation, Eq.~\eqref{eq:ham} is Fourier transformed to momenta $-\pi/a_{y,z}\leq k_{y,z}<\pi/a_{y,z}$, where $a_{y}$ ($a_z$) is the unit cell distance in $y$ ($z$) direction. The Fourier transform of $\Psi_{k_xpq}^\pdag$ is $\Psi_{\bf{k}}^\pdag$, where ${\bf{k}}=(k_x,k_y,k_z)^T$ denotes the 3D momentum. Next, it is useful to perform the gauge transformations $c_{\bf{k}}^{(\rm{B})}\to e^{ik_ya_y/2}\,c_{\bf{k}}^{(\rm{B})}$, $c_{\bf{k}}^{(\rm{C})} \to e^{-ik_za_z/2}\, c_{\bf{k}}^{(\rm{C})}$, $c_{\bf{k}}^{(\rm{D})} \to e^{ik_ya_y/2}\,e^{-ik_za_z/2}\, c_{\bf{k}}^{(\rm{D})}$, and to introduce a pseudospin $\sigma$ within the A, B, and C, D subspaces, acted on by Pauli matrices $\sigma_{x,y,z}$, as well as a pseudospin $\tau$ between these two subspaces. The Hamiltonian can then be cast into the form $H=\sum_{\bf{k}}\Psi_{\bf{k}}^\dagger\,\mathcal{H}_{\bf k}\,\Psi_{\bf{k}}^\pdag$ with

\begin{align}
\mathcal{H}_{\bf{k}}=&v_F k_x\,\sigma_z\tau_z+(t_{y1}+t_{y2})\cos(k_ya_y/2)\,\sigma_x\nonumber
\\&-(t_{y1}-t_{y2})\sin(k_ya_y/2)\,\sigma_y\tau_z\nonumber\\
&+(t_{z1}+t_{z2})\cos(k_za_z/2)\,\tau_x\nonumber\\
&+(t_{z1}-t_{z2})\sin(k_za_z/2)\,\tau_y~.\label{eq:final_ham}
\end{align}
After the canonical transformation $\sigma_y\to\sigma_y\tau_z$, $\sigma_z\to\sigma_z\tau_z$, $\tau_x\to\tau_x\sigma_x$, $\tau_y\to\tau_y\sigma_x$, the diagonalization of the $\tau$-sector yields $H=\sum_{\bf{k}}\Psi_{\bf{k}}^\dagger\,{\rm{diag}}(\mathcal{H}_{{\bf k},+},\mathcal{H}_{{\bf k},-})\,\Psi_{\bf{k}}^\pdag$ with

\begin{align}
\mathcal{H}_{\bf{k},\pm}=&v_F k_x\sigma_z-(t_{y1}- t_{y2})\,\sin(k_ya_y/2)\sigma_y+M_{\pm}(\textbf{k})\sigma_x~,
\end{align}
where the momentum-dependent mass reads

\begin{align}
M_{\pm}(\textbf{k})=&\pm\sqrt{t_{z1}^2+t_{z2}^2+2t_{z1}t_{z2}\cos(k_z a_z)}\nonumber\\
&+(t_{y1}+ t_{y2})\,\cos(k_ya_y/2)~.
\end{align}

\section{Phase diagram of the non-interacting model}\label{sec:non_int_phases}
Despite being described by the same low-energy Hamiltonian as stacked topological insulators, I find that the array of coupled wires can enter a single-surface quantum anomalous Hall (SSQAH) phase which is not present in Ref.~\onlinecite{burkov_balents_layers_11}. This phase is realized for $t_{y1}+t_{y2} < |t_{z1}-t_{z2}|$ and $t_{z1}<t_{z2}$, and has a single quantum Hall layer formed by the the topmost (for $t_{y1}<t_{y2}$) or bottommost (for $t_{y1}>t_{y2}$) layer of wires, see below. Interestingly, the ratio $t_{y1}/t_{y2}$ thus provides an experimental knob (``quantum Hall switch'') selecting the surface on which the single quantum Hall layer, heralded by its gapless edge state, appears.

  \begin{figure}
  \centering
  \includegraphics[width=\columnwidth]{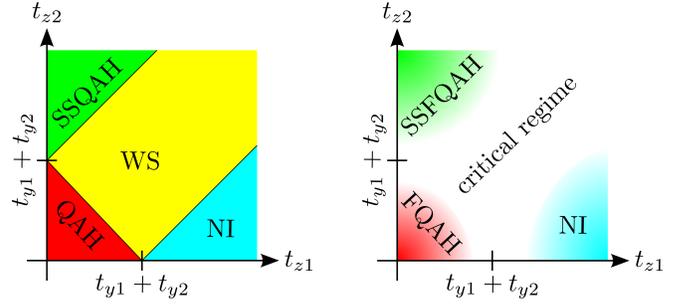}
  \caption{Phase diagram in the non-interacting (left), and interacting (right) case with normal insulating (NI), (fractional) quantum anomalous Hall ((F)QAH), Weyl semimetal (WS), and single-surface (fractional) quantum anomalous Hall (SS(F)QAH) phases. In a lattice model, the filling would be 1/2 (1/6) in the non-interacting (interacting) case.}
  \label{fig:phases}
\end{figure}

All other phases of the array are, however, similar to Ref.~\onlinecite{burkov_balents_layers_11}. For $t_{y1}+t_{y2} < |t_{z1}-t_{z2}|$ and $t_{z1}>t_{z2}$, the system is a normal insulator (NI). If $|t_{z1}-t_{z2}|\leq t_{y1}+t_{y2} \leq t_{z1}+t_{z2}$, the array forms a Weyl semimetal (WS) with two gapless Weyl nodes of opposite chirality at $\textbf{k}_\pm=(0,0,\pi/a_z\pm k_{z0})^T$, where $k_{z0}=\arccos(1-[(t_{y1}+t_{y2})^2-(t_{z1}-t_{z2})^2]/2 t_{z1}t_{z2})/a_z$. Surfaces then have a Fermi arc between the projection of the Weyl nodes, and show a finite Hall conductivity proportional to the distance of the projected Weyl nodes. For $t_{y1}+t_{y2} > t_{z1}+t_{z2}$, finally, the system is in a quantum anomalous Hall (QAH) phase, and exhibits a Hall conductivity of $\sigma_{xy}=e^2/h$ per $(x,y)$ layer of units cells on surfaces with a  normal in the $(x,y)$ plane. The phase diagram of the array of wires is shown in Fig.~\ref{fig:phases}. I have checked that the bulk phase transitions as obtained from the effective low-energy model, and the presence of a Weyl semimetal phase qualitatively agree with a tight-binding calculation that includes all wires and tunnelings (also the subleading tunnelings between wires X, C, and D, as well as Y, A, and B).

\begin{figure}
  \centering
  \includegraphics[width=1.0\columnwidth]{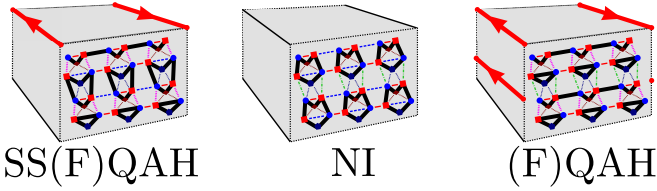}
  \caption{Visualization of the gapped phases for $t_{y1}<t_{y2}$. Thick solid lines depict the dominant couplings along which closed loops and open planes of integer or fractional quantum Hall layers form. Open planes are associated with chiral edge modes illustrated by thick lines with arrows.}
  \label{fig:visu}
\end{figure}

Besides a blueprint for the constructivist engineering of 3D topological states, the coupled-wire construction provides particularly simple visualizations of the bulk gapped phases based on their hierarchy of couplings. The subleading couplings, which compete with the respective dominant coupling, are irrelevant, and the system is adiabatically connected to an array in which only the leading coupling is present. Along the dominant tunnelings, the edge modes of the individual building blocks form closed loops and open planes of quantum Hall layers, see Fig.~\ref{fig:visu}. In the NI phase, all building blocks connect along closed loops, resulting in a full gap. The QAH phase, on the other hand, consists of open quantum Hall planes alternating with closed quantum Hall loops. Since each open plane has a gapless edge mode, the Hall conductivity is indeed $\sigma_{xy}=e^2/h$ per $(x,y)$ layer of units cells. In the SSQAH phase, all building blocks form trivial quantum Hall loops, except for the ones on the topmost or bottommost layer (depending on the ratio $t_{y1}/t_{y2}$). These form a single open quantum Hall plane. The WS, finally, occurs if the competing tunnelings along $y$ and $z$ are of similar strength. The system then enters a critical phase with gapless states of definite momentum (the Weyl nodes), which has no simple real-space picture.

\section{Electron-electron interactions and fractionalization}\label{sec:interactions_ll}
While the non-interacting array allowed the construction and analysis of the interesting SSQAH state, and gives simple physical pictures of the bulk gapped phases, the true power of coupled-wire constructions lies in the description of interacting systems. To tackle these, I linearize the spectrum of each wire $n=A,B,C,D,X,Y$ within a given unit cell around the Fermi points at $\pm k_{1,2,3}$, and decompose the electron operators into right ($R$) and left ($L$) moving modes as $c_n(x)=e^{ik_{FRn}x} R_n(x)+e^{ik_{FLn}x} L_n(x)$, where the respective Fermi momentum is $k_{Frn}$, and $r=R,L$. For $k_1=m k_3$ and $k_2 = (m+1) k_3$ with an integer $m>0$, momentum conservation forbids single-particle tunneling between counterpropagating modes. The combination of electron-electron interactions and tunneling, however, still generates momentum conserving correlated tunnelings between neighboring wires. 

In the $2m$th order of a perturbation theory in a (contact) interaction $V$, the tunneling $t_{A-X}$ between wires $A$ and $X$ generates for instance a term proportional to

\begin{align}
t_{A-X}\,V^{2m}& \left(e^{-i k_{FRX} x}R_X^\dagger\,e^{i k_{FLA}x}L_A^\pdag\right)\nonumber\\
\times&\left(e^{-i k_{FRX} x}R_X^\dagger\,e^{i k_{FLX}x}L_X^\pdag\right)^m\nonumber\\
\times&\left(e^{-i k_{FRA} x}R_A^\dagger\,e^{i k_{FLA}x}L_A^\pdag\right)^m+\rm{H.c.}~.
\end{align}
Connecting different Fermi points, this term trivially conserves energy. Momentum, on the other hand, is conserved for $(m+1)\,k_{FRX}+m\,k_{FRA}=(m+1)\,k_{FLA}+m\,k_{FLX}$, which is fulfilled for the considered $k_1=m k_3$ and $k_2 = (m+1) k_3$. Collecting all leading edge mode scatterings at this order in $V$ that conserve energy and momentum, I find that the bulk of the A-X-B (C-Y-D) building blocks can be fully gapped by

\begin{subequations}
\begin{align}
R_{A(D)}^\dagger{}^mR_{X(Y)}^\dagger{}^{m+1}L_{X(Y)}^\pdag{}^mL_{A(D)}^\pdag{}^{m+1}+\rm{H.c.}~,\\
R_{X(Y)}^\dagger{}^mR_{B(C)}^\dagger{}^{m+1}L_{B(C)}^\pdag{}^mL_{X(Y)}^\pdag{}^{m+1}+\rm{H.c.}~.
\end{align}\label{eq:block_couplings}
\end{subequations}
The leading couplings between the edge modes of the building blocks inside a unit cell, on the other hand, are

\begin{subequations}
\begin{align}
R_B^\dagger{}^mR_A^\dagger{}^{m+1}L_A^\pdag{}^mL_B^\pdag{}^{m+1}+\rm{H.c.}~,\label{eq:coupling1}\\
R_C^\dagger{}^mR_D^\dagger{}^{m+1}L_D^\pdag{}^mL_C^\pdag{}^{m+1}+\rm{H.c.}~,\label{eq:coupling2}\\
R_C^\dagger{}^mR_A^\dagger{}^{m+1}L_A^\pdag{}^mL_C^\pdag{}^{m+1}+\rm{H.c.}~,\label{eq:coupling3}\\
R_B^\dagger{}^mR_D^\dagger{}^{m+1}L_D^\pdag{}^mL_B^\pdag{}^{m+1}+\rm{H.c.}\label{eq:coupling4}
\end{align}\label{eq:edge_couplings}
\end{subequations}
Couplings between different unit cells will be treated in section \ref{subsec:coupling}.

\subsection{Luttinger liquid description of an individual building blocks}\label{subsec:blocks}
To understand how these couplings can lead to gapped Laughlin states at filling $\nu=1/(2m+1)$ within each building block, it is most convenient to switch to a bosonized language.\cite{giamarchi_book} Since the A-X-B and C-Y-D blocks can be treated in an analogous fashion, I only detail the description of the former. The relevant low-energy degrees of freedom are thus the chiral modes close to the Fermi points at momentum $k_x=\pm k_{1,2,3}$ in wires $n=A,X,B$, which are bosonized as
\begin{align}
  r_{n}^{}(x)=
  \frac{U_{rn}^{}}{\sqrt{2\pi\alpha}}\,e^{-i\Phi_{rn}(x)},\label{eq:bos_form}
\end{align}
where $U_{rn}^{}$ is a Klein factor, while $\alpha^{-1}$ denotes a high momentum cutoff. The bosonized fields obey
\begin{align}
  \bigl[\Phi_{rn}(x),\Phi_{r'n'}(x')\bigr] =\delta_{rr'}\delta_{nn'}\,i\pi r\,\text{sgn}(x-x').
\end{align}
It is furthermore helpful to also introduce the fields

\begin{subequations}
\begin{align}
\widetilde{\Phi}_{Rn}^{(m)} &= (m+1)\Phi_{R n}-m\Phi_{Ln}~,\\
\widetilde{\Phi}_{Ln}^{(m)} &= (m+1)\Phi_{L n}-m\Phi_{R k}~.
\end{align}\label{eq:basis_trafo}
\end{subequations}
These obey the commutator

\begin{align}
[\widetilde{\Phi}_{rn}^{(m)}(x),\widetilde{\Phi}_{r'n'}^{(m)}(x')]=\delta_{rr'}\delta_{nn'}(2m+1)\,i\pi r\,\text{sgn}(x-x')~.\label{eq:supp_comm}
\end{align}
Dropping the Klein factors for notational convenience, Eq.~\eqref{eq:block_couplings} translates to sine-Gordon terms

\begin{subequations}
\begin{align}
\cos\left(\widetilde{\Phi}_{RX}^{(m)}-\widetilde{\Phi}_{LA}^{(m)}\right)~,\\
\cos\left(\widetilde{\Phi}_{RB}^{(m)}-\widetilde{\Phi}_{LX}^{(m)}\right)~.
\end{align}\label{eq:supp_ham2}
\end{subequations}
To show that these couplings can stabilize a three-wire wide quantum Hall state, I follow Ref.~\onlinecite{kane_02}, and note that the argument of each sine-Gordon term commutes with itself at different positions. These cosines can thus order individually by pinning their arguments to one of their minima. In addition, they can order simultaneously, since their arguments also commute between each other. Because one can always find a Hamiltonian that renders the different sine-Gordon terms relevant in the renormalization group (RG) sense,\cite{kane_02} it is always possible to reach this situation at low energies. The resulting bulk gapped state has  quasiparticle excitations of charge $e/(2m+1)$, as can be inferred from kinks in the cosines,\cite{kane_02,charge_remark} and chiral edge modes $\widetilde{\Phi}_{RA}$ and $\widetilde{\Phi}_{LB}$ that obey the commutator given in Eq.~\eqref{eq:supp_comm}, as expected for a Laughlin state at filling $\nu=1/(2m+1)$.\cite{kane_02} It has furthermore been shown that also the Chern-Simons term associated with a 2D (fractional) quantum Hall effect is contained in the coupled-wire construction used for the individual building blocks.\cite{santos_15} All of these signatures combined allow to positively identify the gapped phase resulting from the coupling in Eq.~\eqref{eq:supp_ham2} as an integer (for $m=0$) or fractional (for integer $m>0$) quantum Hall state. The integer $m$, which technically relates to the order in perturbation theory in the interaction $V$ that generates the required  sine-Gordon term, thus physically indicates the effective filling fraction $\nu=1/(2m+1)$ of the individual building blocks. 

In a larger and isolated building block, where there is neither overlap nor direct tunneling between them, the edge modes $\widetilde{\Phi}_{RA}^{(m)}$ and $\widetilde{\Phi}_{LB}^{(m)}$ would be gapless (they simply do not appear in the sine-Gordon part of the Hamiltonian, and therefore only carry a kinetic energy $\sim \pm v_F k_x$). In the narrow building blocks considered here, both a finite overlap through the bulk (the wire X), and a direct tunneling between the edge modes exist, such that the edge modes are gapped. The leading edge mode tunneling  within the $A-X-B$ building block that is compatible with the bulk sine-Gordon terms in Eq.~\eqref{eq:supp_ham2} is given in Eq.~\eqref{eq:coupling1}. Its bosonized form

\begin{align}
\cos\left(\widetilde{\Phi}_{RA}^{(m)}-\widetilde{\Phi}_{LB}^{(m)}\right)
\end{align}
shows that this term can order, and is indeed compatible with the bulk cosines: its argument commutes with itself at different positions, and with the arguments of the cosines given in Eq.~\eqref{eq:supp_ham2}. The couplings between the edges of the the $C-Y-D$ block and between the two building blocks of a unit cell, given in Eq.~\eqref{eq:coupling2}-\eqref{eq:coupling4}, have the bosonized expressions

\begin{subequations}
\begin{align}
\cos\left(\widetilde{\Phi}_{RD}^{(m)}-\widetilde{\Phi}_{LC}^{(m)}\right)~,\\
\cos\left(\widetilde{\Phi}_{RA}^{(m)}-\widetilde{\Phi}_{LC}^{(m)}\right)~,\\
\cos\left(\widetilde{\Phi}_{RD}^{(m)}-\widetilde{\Phi}_{LB}^{(m)}\right)~.
\end{align}
\end{subequations}
The arguments of the individual cosines commute with themselves at different positions, and with the arguments of the bulk sine-Gordon terms. The arguments of different sine-Gordon terms involving the same edge mode do, however, not commute, such that these edge mode couplings cannot order simultaneously.

\subsection{Luttinger liquid description of the three-dimensional system}\label{subsec:coupling}
As in the non-interacting case, I furthermore subject the edge modes of the building blocks in the 3D system of coupled quantum wires to inter block couplings. I find that the leading edge mode couplings compatible with the bulk couplings of the individual building blocks in Eq.~\eqref{eq:block_couplings} read
\begin{subequations}
\begin{align}
R_{Bpq}^\dagger{}^mR_{Ap'q'}^\dagger{}^{m+1}L_{Ap 'q'}^\pdag{}^mL_{Bpq}^\pdag{}^{m+1}+\rm{H.c.}~,\\
R_{Cpq}^\dagger{}^mR_{Dp'q'}^\dagger{}^{m+1}L_{Dp'q'}^\pdag{}^mL_{Cpq}^\pdag{}^{m+1}+\rm{H.c.}~,\\
R_{Cpq}^\dagger{}^mR_{Ap'q'}^\dagger{}^{m+1}L_{Ap'q'}^\pdag{}^mL_{Cpq}^\pdag{}^{m+1}+\rm{H.c.}~,\\
R_{Bpq}^\dagger{}^mR_{Dp'q'}^\dagger{}^{m+1}L_{Dp'q'}^\pdag{}^mL_{Bpq}^\pdag{}^{m+1}+\rm{H.c.}~,
\end{align}\label{eq:supp_edge_couplings}
\end{subequations}
where $(pq)$ and $(p'q')$ are the $(xy)$-indices of neighboring unit cells. Like the intra unit cell couplings, these terms can be generated by a perturbation theory in an interaction $V$, and inter wire tunneling (the intra unit cell couplings in Eq.~\eqref{eq:edge_couplings} can be recovered by setting $p=p'$ and $q=q'$). The couplings in Eq.~\eqref{eq:supp_edge_couplings} conserve energy and momentum for the considered case of $k_1=m\, k_3$ and $k_2 = (m+1)\, k_3$. Their bosonized form is

\begin{subequations}
\begin{align}
\cos\left(\widetilde{\Phi}_{RAp'q'}^{(m)}-\widetilde{\Phi}_{LBpq}^{(m)}\right)~,\\
\cos\left(\widetilde{\Phi}_{RDp'q'}^{(m)}-\widetilde{\Phi}_{LCpq}^{(m)}\right)~,\\
\cos\left(\widetilde{\Phi}_{RAp'q'}^{(m)}-\widetilde{\Phi}_{LCpq}^{(m)}\right)~,\\
\cos\left(\widetilde{\Phi}_{RDp'q'}^{(m)}-\widetilde{\Phi}_{LBpq}^{(m)}\right)~,
\end{align}\label{eq:supp_edge_couplings_bos}
\end{subequations}
where $\widetilde{\Phi}_{rnpq}^{(m)}$ is the bosonized field associated with the edge mode $r=R,L$ in wire $n=A,B,C,D$ of unit cell $(pq)$ defined in analogy to Eq.~\eqref{eq:basis_trafo} as

\begin{subequations}
\begin{align}
\widetilde{\Phi}_{Rnpq}^{(m)} &= (m+1)\Phi_{R npq}-m\Phi_{Lnpq}~,\\
\widetilde{\Phi}_{Lnpq}^{(m)} &= (m+1)\Phi_{L npq}-m\Phi_{R kpq}~.
\end{align}\label{eq:basis_trafo_app}
\end{subequations}
These fields obey the commutator

\begin{align}
[\widetilde{\Phi}_{rnpq}^{(m)}(x),\widetilde{\Phi}_{r'n'p'q'}^{(m)}(x')]=&\delta_{rr'}\delta_{nn'}\delta_{pp'}\delta_{qq'}\,(2m+1)\nonumber\\&\times i\pi r\,\text{sgn}(x-x')~.\label{eq:supp_comm_2}
\end{align}
The argument of each sine-Gordon term commutes with itself at different positions, and can thus order. Like within a unit cell, the arguments of different sine-Gordon terms involving the same edge mode do not commute, and these sine-Gordon terms cannot order simultaneously. They do, however, commute with the arguments of the sine-Gordon terms stabilizing the integer and fractional quantum Hall states within each building block. The couplings in Eq.~\eqref{eq:supp_edge_couplings_bos} are thus a set of competing edge mode couplings that leave the bulk gaps of the individual building blocks untouched.

\section{Phase diagram of the interaction model}\label{sec:interact_phases}
Within each A-X-B and C-Y-D building block, where all sine-Gordon terms can order simultaneously, a bulk gap results from the pinning of the arguments of the cosines. For the competing edge mode couplings, however, the situation is more delicate: not all sine-Gordon terms can pin their arguments simultaneously. If the couplings have a clear hierarchy in strength, the leading sine-Gordon term orders, while the others are suppressed. The system is then in a phase that is adiabatically connected to the array being only subject to the leading coupling, while all other couplings vanish. If there is no clear hierarchy, the different sine-Gordon terms compete, and there is no simple description in terms of pinned fields.

\subsection{Clear hierarchy of couplings}
Like in the non-interacting case, I take the couplings of the inner and outer wires in each building block to be the largest correlated tunnelings, which puts each building block in a fractional quantum Hall state. To analyze the effect of the competing edge mode couplings, I study an array with the same relative coupling strengths as before (alternating couplings along $y$, shifted by one edge mode in the next $(x,y)$ layer, and alternating couplings along $z$). If the intra unit cell couplings \eqref{eq:coupling3} and \eqref{eq:coupling4} generalizing the hoppings $t_{z1}$ in the fractional case are much stronger than the other edge mode couplings, the A-X-B and C-Y-D building blocks pair up within the unit cells, and the array is in a normal insulating state. To see this, I use that the array is then adiabatically connected to a system where the edge modes are only connected by this dominant hopping. As shown in the middle panel of Fig.~\ref{fig:visu}, the fractional quantum Hall building blocks then form small loops, which indeed corresponds to a normal insulating state. If, however, the edge modes dominantly pair up along $z$ between neighboring unit cells, a single-surface fractional quantum anomalous Hall (SSFQAH) phase with a single fractional quantum Hall layer on either the top or bottom surface is realized, as can be seen in the left panel of Fig.~\ref{fig:visu}. If, finally, one of the couplings along $y$ is strongest, I obtain a fractional quantum anomalous Hall (FQAH) phase analog to the above QAH phase, which has one open fractional quantum Hall plane per $(x,y)$ layer of unit cells, see right panel of Fig.~\ref{fig:visu}. This phase shows a fractional Hall conductivity of $\sigma_{xy}=\nu\,e^2/h$ per $(x,y)$ layer of units cells, where $\nu=1/(2m+1)$. All of these bulk gapped phases are somewhat similar to weak topological insulators in that the individual fractional quantum Hall loops and planes support gapped quasiparticles of fractional charge and statistics confined to their respective 2D subsystem. The phase diagram of the interacting array in Fig.~\ref{fig:phases} locates the discussed phases in the respective limits of the parameter space.

\subsection{Solution of Luther-Emery type at special points in the parameter space}\label{sec:special_points}
Most interesting is the fate of the Weyl semimetal. I argued that this phase emerges in the non-interacting case when several competing tunneling couplings are of comparable strength. While the Weyl semimetal phase can be identified straightforwardly in a fermionic language, as was done in Sec.~\ref{sec:non_int_phases}, it corresponds to a complicated regime of competing sine-Gordon terms of comparable strength in the Luttinger liquid description, see Sec.~\ref{subsec:coupling} with $m=0$. In the interacting array, I find that the same competition can also exist between the now more complicated sine-Gordon terms. The striking mathematical analogy of the Weyl semimetal regime and the regime of competing edge mode couplings with comparable strength in the interacting case hints at the possible existence of an extended critical phase separating the bulk gapped fractional phases. Some support for the generic existence of critical phases in 3D comes from the superconducting version of the non-interacting Hamiltonian in Eq.~\eqref{eq:ham}, which has a gapless superconducting Weyl phase\cite{meng_12} similar to $^3$He-A.\cite{volovik_book} Also spin liquids can exhibit a Weyl phase.\cite{hermanns_15} 

As a first step, it is instructive to analyze the special points in parameter space where only one of the competing sine-Gordon terms is present. For these situations, an exact solution of the low-energy Hamiltonian in the spirit of a Luther-Emery point is possible for special choices of the density-density interactions. For concreteness, let me take only the coupling $t_{z1}^{(m)}$ generalizing $t_{z_1}$ to be finite. The Hamiltonian then reads

\begin{align}
&H=H_{\rm quad}+H_{t_{z1}}^{(m)},\\
&H_{t_{z1}}^{(m)}=\sum_{pq}\int dx\frac{t_{z1}^{(m)}}{\pi\alpha}\cos\left(\widetilde{\Phi}_{RApq}^{(m)}(x)-\widetilde{\Phi}_{LCpq}^{(m)}(x)\right)\nonumber\\
&+\sum_{pq}\int dx\frac{t_{z1}^{(m)}}{\pi\alpha}\cos\left(\widetilde{\Phi}_{LBpq}^{(m)}(x)-\widetilde{\Phi}_{RDpq}^{(m)}(x)\right)
\end{align}
Here, $H_{\rm quad}$ contains the gapless motion along the $x$ direction, and density-density interactions quadratic in the bosonized fields. $H_{\rm quad}$ is thus composed of terms proportional to $\int dx\, (\partial_x \widetilde{\Phi}_{rnpq}^{(m)}) (\partial_x \widetilde{\Phi}_{r'n'p'q'}^{(m)})$.\cite{giamarchi_book}
I now  introduce the new fields

\begin{subequations}
\begin{align}
\bar{\Phi}_{RApq}^{(m)}(x)&=\frac{m+1}{2m+1}\widetilde{\Phi}_{RApq}^{(m)}(x)-\frac{m}{2m+1} \widetilde{\Phi}_{LCpq}^{(m)}(x)~,\\
\bar{\Phi}_{LCpq}^{(m)}(x)&=\frac{m+1}{2m+1}\widetilde{\Phi}_{LCpq}^{(m)}(x)-\frac{m}{2m+1} \widetilde{\Phi}_{RApq}^{(m)}(x)~,\\
\bar{\Phi}_{LBpq}^{(m)}(x)&=\frac{m+1}{2m+1}\widetilde{\Phi}_{LBpq}^{(m)}(x)-\frac{m}{2m+1} \widetilde{\Phi}_{RDpq}^{(m)}(x)~,\\
\bar{\Phi}_{RApq}^{(m)}(x)&=\frac{m+1}{2m+1}\widetilde{\Phi}_{RDpq}^{(m)}(x)-\frac{m}{2m+1} \widetilde{\Phi}_{LBpq}^{(m)}(x)~.
\end{align}\label{eq:full_trafo}
\end{subequations}
These obey the canonical commutator

\begin{align}
[\bar{\Phi}_{rnpq}^{(m)}(x),\bar{\Phi}_{r'n'p'q'}^{(m)}(x')]=&\delta_{rr'}\delta_{nn'}\delta_{pp'}\delta_{qq'}\, i\pi r\,\text{sgn}(x-x')~.\label{eq:supp_comm_3}
\end{align}
Rewriting the Hamiltonian in terms of these new degrees of freedom yields

\begin{align}
&H=\bar{H}_{\rm quad}+\bar{H}_{t_{z1}}^{(m)},\\
&\bar{H}_{t_{z1}}^{(m)}=\sum_{pq}\int dx\frac{t_{z1}^{(m)}}{\pi\alpha}\cos\left(\bar{\Phi}_{RApq}^{(m)}(x)-\bar{\Phi}_{LCpq}^{(m)}(x)\right)\nonumber\\
&+\sum_{pq}\int dx\frac{t_{z1}^{(m)}}{\pi\alpha}\cos\left(\bar{\Phi}_{LBpq}^{(m)}(x)-\bar{\Phi}_{RDpq}^{(m)}(x)\right)~,
\end{align}
where $\bar{H}_{\rm quad}$ is still composed of products of linear derivatives of the new bosonized fields $\bar{\Phi}_{rnpq}^{(m)}$. Because the fields $\bar{\Phi}_{rnpq}^{(m)}$ obey the canonical commutator of the bosonized fields describing chiral fermions, I can now rewrite the Hamiltonian as

\begin{align}
&H=\bar{H}_{0}^{(m)}+\bar{H}_{\rm int},\label{eq:supp_ham_bos_full_special}\\
&\bar{H}_{0}^{(m)}=\int dx\sum_{p,q}\bar{\Psi}_{pq}^{(m)}{}^\dagger(x)\,\mathcal{H}^{(m)}_0(x)\,\bar{\Psi}_{pq}^{(m)}(x)^\pdag~,\\
&\mathcal{H}^{(m)}_0(x)=\begin{pmatrix}-iu \partial_x&0&t_{z_1}^{(m)}&0\\0&iu \partial_x&0&t_{z_1}^{(m)}\\t_{z_1}^{(m)}&0&iu \partial_x&0\\0&t_{z_1}^{(m)}&0&-iu \partial_x\end{pmatrix}~,\label{eq:h_cos}\\
&\bar{\Psi}_{pq}^{(m)} =\frac{1}{\sqrt{2\pi\alpha}} (e^{-i\bar{\Phi}_{RApq}^{(m)}},e^{-i\bar{\Phi}_{LBpq}^{(m)}},e^{-i\bar{\Phi}_{LCpq}^{(m)}},e^{-i\bar{\Phi}_{RDpq}^{(m)}})^T~,\label{eq:new_op_special}
\end{align}
where $\bar{H}_{\rm int}$ describes only the density-density interactions between the different modes, and where $u$ is the velocity associated with the fields $\bar{\Phi}_{rnpq}^{(m)}$ (which is for simplicity assumed to be identical for all modes). I can now invert the bosonization procedure for the components of $\bar{\Psi}_{pq}^{(m)}$ for all $m$ using the refermionization procedure.\cite{giamarchi_book,delf_schoeller_review} Diagonalizing  the refermionized version of $\bar{H}_{0}^{(m)}$ as in Sec.~\ref{sec:non_int_array} then allows to calculate its spectrum for all $m$. As follows from Sec.~\ref{sec:non_int_array} upon setting all tunnel couplings except $t_{z_1}$ to zero, the system is in a normal insulating state for $t_{z_1}^{(m)}\neq0$, and has a critical point for $t_{z_1}^{(m)}=0$.

It is interesting to also determine the charge associated with the operator $e^{i\bar{\Phi}_{rnpq}^{(m)}}$, which can be inferred from the bosonized expression of the total charge density operator,

\begin{align}
\rho_{\rm tot}&=\sum_{pq}\sum_{n=A,B,C,D}\frac{-e}{2\pi}\partial_x(\Phi_{Rnpq} -\Phi_{Lnpq})\nonumber\\
&=\sum_{pq}\sum_{n=A,B,C,D}\frac{-e}{2\pi(2m+1)}\partial_x(\widetilde{\Phi}_{Rnpq} -\widetilde{\Phi}_{Lnpq})\nonumber\\
&=\sum_{pq}\sum_{n=A,B,C,D}\frac{-e}{2\pi(2m+1)}\partial_x(\bar{\Phi}_{Rnpq} -\bar{\Phi}_{Lnpq})~,\label{eq:charge}
\end{align}
where $e$ denotes the charge of an electron. Commuting $\rho_{\rm tot}$ and $e^{i\bar{\Phi}_{rnpq}^{(m)}}$, I find that these latter exponentials create quasiparticles of charge $e/(2m+1)$. This finding agrees with the observation that $2\pi$-kinks (the smallest possible kinks) in any of the bulk sine-Gordon terms given in Eq.~\eqref{eq:supp_edge_couplings_bos} carry charge $e/(2m+1)$,\cite{charge_remark}  and that quasiparticle operators add local excitations above the ground state. Indeed, kinks in the bulk sine-Gordon terms precisely correspond to local excitations above the ground state. Since the analysis performed here for only $t_{z1}^{(m)}$ being finite can be directly transposed to any other coupling, I find that if only one of the competing tunnel couplings is present, the system can be described in terms of interacting fermionic quasiparticles of fractional charge $e/(2m+1)$ that have a critical point, at which they disperse linearly in $x$ direction. In the spirit of a Luther-Emery point, one can now consider fine-tuning the density-density interactions in the system to reach $\bar{H}_{\rm int}=0$. At this point, the non-interacting spectrum of $\bar{H}_{0}^{(m)}$ is the true spectrum of the system, and the strongly interacting electron system has  been mapped exactly to free fermions of fractional charge. As for a one-dimensional Mott insulator at half-filling with a Luttinger liquid parameter $K=1/2$, these emerging fermions are complicated objects that are not to be confused with the original electrons in the system.\cite{giamarchi_book}

\subsection{Competing couplings of comparable strength}\label{subsec:comp_strength}
When more than one of the competing sine-Gordon couplings is present, the transformation in Eq.~\eqref{eq:full_trafo} is not sufficient to rewrite the Hamiltonian in terms of (free) fermions anymore. For the non-interacting case $m=0$, I showed in Sec.~\ref{sec:non_int_phases} that the effect of the additional couplings is to turn the critical point associated with $t_{z1}=0$ considered in the last subsection into an extended critical phase, the Weyl semimetal. Put in simple terms, the additional couplings  allow the gapless quasiparticles to move in 3D. For finite $m$, one may suspect that the critical point associated with $t_{z_1}^{(m)}=0$ considered in the last subsection also turns into an extended critical 3D phase when the additional couplings are turned on. Similar to a discussion supporting fractional topological insulators in Ref.~\onlinecite{sagi_preprint}, this scenario can be supported by rewriting the Hamiltonian as

\begin{align}
&H=H_{\rm quad}+H_{\rm cos}~,\\
&H_{\rm cos}=\int dx\sum_{p,q,p'q'}\Psi_{pq}^{(m)}{}^\dagger(x)\,\mathcal{H}^{pp'qq'}(x)\,\Psi_{p'q'}^{(m)}(x)^\pdag~,\label{eq:supp_ham_bos_full}\\
&\mathcal{H}^{pp'qq'}=\begin{pmatrix}\mathcal{H}_{11}^{pp'qq'}(x)&\mathcal{H}_{12}^{pp'qq'}(x)\\\mathcal{H}_{21}^{pp'qq'}(x)&\mathcal{H}_{22}^{pp'qq'}(x)\end{pmatrix}~,\label{eq:h_cos}\\
&\Psi_{pq}^{(m)} =\frac{1}{\sqrt{2\pi\alpha}} (e^{-i\widetilde{\Phi}_{RApq}^{(m)}},e^{-i\widetilde{\Phi}_{LBpq}^{(m)}},e^{-i\widetilde{\Phi}_{LCpq}^{(m)}},e^{-i\widetilde{\Phi}_{RDpq}^{(m)}})^T~.\label{eq:new_op}
\end{align}
Here, the $(2\times2)$ matrices $\mathcal{H}_{ij}^{pp'qq'}(x)$ contain the prefactors of the sine-Gordon terms. By adjusting the tunnel couplings and interaction strengths, $\mathcal{H}^{pp'qq'}$ can be tuned to be of the same form for all integers $m$, including the non-interacting case $m=0$. Consequently, the same transformations that diagonalize the Hamiltonian matrix $\mathcal{H}^{pp'qq'}$ for $m=0$ also do so for positive integers $m$. These transformations follow from Eq.~\eqref{eq:ham} and its discussion in Sec.~\ref{sec:non_int_array} upon setting $v_Fk_x\to0$, and replacing the non-interacting tunnelings $t_{y,z,1,2}$ by the prefactors of the corresponding sine-Gordon terms. As one important step in this diagonalization, I highlight the Fourier transformation along $y$ and $z$, which yields $H_{\rm cos}=\sum_{k_yk_z}\int dx\,\Psi_{k_yk_z}^{(m)}{}^\dagger(x)\,\mathcal{H}_{k_yk_z}(x)\,\Psi_{k_yk_z}^{(m)}(x)$ with $\Psi_{k_yk_z}^{(m)}(x) = \frac{1}{\sqrt{N_yN_z}}\sum_{pq} e^{-i (k_y a_y p+k_z a_z q)} \Psi_{pq}^{(m)}(x)$, where $N_{y(z)}$ is the number of unit cells in $y$ ($z$) direction. 

Within the region of the phase diagram hosting the Weyl semimetal in the non-interacting case $m=0$, also the interacting array has excitations $\Psi_{k_yk_z}^{(m)}(x)$ associated with momentum $k_y=0$ and $k_z=\pi/a_z\pm k_{z0}$ that do not appear in the Hamiltonian $H_{\rm cos}$, but only in $H_{\rm quad}$ (which accounts for the possibly interaction-renormalized kinetic energy along the $x$ direction). For $m=0$, these are precisely the gapless electronic quasiparticles at the Weyl nodes. Outside the parameter regime hosting the Weyl semimetal in the non-interacting case, the Hamiltonian matrix  $\mathcal{H}_{k_yk_z}(x)$ does not vanishes for any $(k_{y},k_{z})$. As a consequence, all excitations create kinks of non-trivial energy in the sine-Gordon Hamiltonian $H_{\rm cos}$, and are thus gapped. This suggests that the topology of the phase diagram is the same in the interacting and non-interacting cases. 

As discussed in the last subsection, the fermionic character of the operators $\Psi_{pq}^{(0)}$ for $m=0$ can most easily be derived by refermionizing its components $e^{-i\widetilde{\Phi}_{rnpq}^{(0)}}$. For $m>0$, however, the fields $\widetilde{\Phi}_{rnpq}^{(m)}$ do not obey the same commutator as the non-interacting fields $\widetilde{\Phi}_{rnpq}^{(0)}$. This implies that the refermionization procedure does not carry over to the interacting case, and that the simple operators $e^{-i\widetilde{\Phi}_{rnpq}^{(m)}}$ do, for general $m$, not satisfy the canonical fermionic commutation relations. This is in agreement with the observation that the operators 

\begin{align}
e^{i\widetilde{\Phi}_{rnpq}^{(m)}}& =\left(e^{i\widetilde{\Phi}_{rnpq}^{(0)}}\right)^{m+1}\,\left(e^{-i\widetilde{\Phi}_{(-r)npq}^{(0)}}\right)^{m}\label{eq:decompose}
\end{align}
correspond to products of $2m+1$ fermionic operators, rather than a single canonical fermion. The operators $\Psi_{k_yk_z}^{(m)}(x)$ with $m>0$ thus do not correspond to individual 3D quasiparticle excitations of the array of wires. Further extensions of the coupled wire construction are needed to determine if also the 3D quasiparticles at finite $m$ have linearly dispersing band touchings similar to Weyl nodes. Similarly, the properties of the interacting analog of the Weyl semimetal phase, such as its response to electro-magnetic fields, cannot be resolved by the present analysis based on a language of competing sine-Gordon terms.

Despite the fact that the present bosonized calculation does not allow to access the true quasiparticles in the generalized Weyl semimetal regime for finite $m$, I judge the striking analogy between the interacting and non-interacting cases, and (for all $m$) the existence of operators that do not appear in $H_{\rm cos}$ in the regime hosting the Weyl semimetal for $m=0$, and only in this regime, to strongly support the existence of an extended critical phase separating the bulk gapped fractional phases emerging from the critical points discussed in Sec.~\ref{sec:special_points}, similar to the Weyl semimetal emerging from the analogous points point for $m=0$. While one option for a possible critical phase would be a standard Weyl semimetal, I note that transitions between the bulk gapped phases correspond to a rearrangement of only the edge states of the individual fractional quantum Hall building blocks. Their bulk gaps in the X and Y wires, however, never close. I thus speculate that the quantum phase transitions of the interacting array, and the possible gapless critical phase inherit some of the exotic properties of fractional quantum Hall edge modes. A critical phase could for instance be composed of fractionally charged fermionic 3D quasiparticles (somewhat related ideas are discussed in Refs.~\onlinecite{levin_09,wang_15_fract_ws}). This scenario is supported by the quasiparticles at the Luther-Emery-type points considered in Sec.~\ref{sec:special_points}.

\section{Summary and conclusions}\label{sec:summary}
In this work, the coupled-wire approach has been adapted for the analysis of interacting topological phases in 3D. More precisely, I analyzed the 3D system resulting from connecting narrow 2D integer and fractional quantum Hall building blocks along different directions. The coupled-wire language is thereby used for the microscopic description of the individual building blocks, see Sec.~\ref{sec:blocks}, and the inter block couplings. For the non-interacting array of wires detailed in Sec.~\ref{sec:non_int_array}, I found that the phase diagram includes a normal insulating, a quantum anomalous Hall, a Weyl semimetal, and a single-surface quantum anomalous Hall phase, see Sec.~\ref{sec:non_int_phases}. The surface on which this latter phase appears can be controlled by the ratio of two tunnel couplings. The coupled-wire language offers simple real space visualizations of the bulk gapped phases in terms of closed loops and open planes of 2D quantum Hall subsystems. With electron-electron interactions, I found that the array can be in analogous bulk gapped phases composed of closed loops and open planes of fractional quantum Hall states, see Sec.~\ref{sec:interactions_ll} and Sec.~\ref{sec:interact_phases}. Some of these phases have topologically protected fractional edge states. In analogy to the non-interacting Weyl semimetal, the bulk gapped phases are separated by a regime in which certain excitation operators do not appear in the sine-Gordon part of the Hamiltonian. This supports the existence of an exotic critical phase similar to a Weyl phase. The phase transitions of the array are of exotic nature, and correspond to a rearrangement of fractional quantum Hall edge modes. If only a single sine-Gordon term is present, the exotic nature of the phases and phase transitions could be revealed by an exact mapping, which identified the quasiparticles at special Luther-Emery points as fermions of fractional charge $e/(2m+1)$. The properties of the critical regime outside these special points will be addressed in future work. Other interesting directions include the use of chiral spin liquids, topological superconductors, and states with a more complex edge structure (such as Moore-Read states) as 2D building blocks, as well as the coupled-wire analysis of non-Abelian 3D phases with string-like excitations. Experimentally, the proposed array could be realized with cold atoms, in which the necessary ingredients (fermions subject to spin-orbit coupling and magnetic field,\cite{soi_coldatoms_1,soi_coldatoms_2} complex lattices,\cite{cold_atoms_lattices} and superlattices hosting for instance even resonating valence-bond states\cite{cold_atoms_unit_cell}) have been demonstrated.

\textit{Note added.} Recetly, a related preprint discussing coupled-wire constructions of 3D integer and fractional topological insulators appeared.\cite{sagi_preprint}

\begin{acknowledgments}
This work has been supported by the Helmholtz association through VI-521, and the DFG through SFB 1143. The author acknowledges stimulating discussions with E. Sela, S. Rachel, M. Vojta, A. Grushin, J. Bardarson and K. Shtengel.
\end{acknowledgments}



\begin{thebibliography}{99}

\bibitem{tsui_fqhe}
D. C. Tsui, H. L. Stormer, and A. C. Gossard, Phys. Rev. Lett. \textbf{48}, 1559 (1982).

\bibitem{laughlin_83}
R. B. Laughlin, Phys. Rev. Lett. \textbf{50}, 1395 (1983).

\bibitem{pesin_10}
D. Pesin and L. Balents, Nat. Phys. \textbf{6}, 376 (2010).

\bibitem{wk_10}
W. Witczak-Krempa, T. P. Choy, and Y. B. Kim, Phys. Rev. B \textbf{82}, 165122 (2010).


\bibitem{scheurer_15}
M. S. Scheurer, S. Rachel, and P. P. Orth, Sci. Rep. \textbf{5}, 8386 (2015).

\bibitem{maciejko_10}
J. Maciejko, X.-L. Qi, A. Karch, and S.-C. Zhang, Phys. Rev. Lett. \textbf{105}, 246809 (2010).

\bibitem{swingle_11}
B. Swingle, M. Barkeshli, J. McGreevy, and T. Senthil, Phys. Rev. B \textbf{83}, 195139 (2011).

\bibitem{maciejko_12}
J. Maciejko, X.-L. Qi, A. Karch, and S.-C. Zhang, Phys. Rev. B \textbf{86}, 235128 (2012).

\bibitem{lee_12}
S. Bhattacharjee, Y. B. Kim, S.-S. Lee, and D.-H. Lee, Phys. Rev. B \textbf{85}, 224428 (2012).

\bibitem{castelnovo_08}
C. Castelnovo and C. Chamon, Phys. Rev. B \textbf{78}, 155120 (2008).

\bibitem{levin_11}
M. Levin, F. J. Burnell, M. Koch-Janusz, and A. Stern, Phys. Rev. B \textbf{84}, 235145 (2011).

\bibitem{kitaev_06}
A. Kitaev, Annals of Physics \textbf{321}, 2 (2006).

\bibitem{si_08}
T.  Si and Y. Yu, Nuclear Physics B \textbf{803}, 428 (2008).

\bibitem{mandal_09}
S. Mandal and N. Surendran, Phys. Rev. B \textbf{79}, 024426 (2009).

\bibitem{ryu_09}
S. Ryu, Phys. Rev. B \textbf{79}, 075124 (2009).

\bibitem{wu_09}
C. Wu, D. Arovas, and H.-H. Hung, Phys. Rev. B \textbf{79}, 134427 (2009).

\bibitem{hermanns_14}
M. Hermanns and S. Trebst, Phys. Rev. B \textbf{89}, 235102 (2014).

\bibitem{hermanns_15}
M. Hermanns, K. O'Brien, and S. Trebst,  Phys. Rev. Lett. \textbf{114}, 157202 (2015).

\bibitem{castelnovo_07}
C. Castelnovo, R. Moessner, and S. Sondhi, Nature \textbf{451}, 42 (2007).

\bibitem{morris_09}
D. J. P. Morris \textit{et al.}, Science \textbf{326}, 411 (2009).

\bibitem{fennel_09}
T. Fennell, P. P. Deen, A. R. Wildes, K. Schmalzl, D. Prabhakaran, A. T. Boothroyd, R. J. Aldus, D. F. McMorrow, and S. T. Bramwell, Science \textbf{326}, 415 (2009).

\bibitem{bramwell_09}
S. T. Bramwell, S. R. Giblin, S. Calder, R. Aldus, D. Prabhakaran, and T. Fennell, Nature \textbf{461}, 956 (2009).

\bibitem{balents_96}
L. Balents and M. P. A. Fisher, Phys. Rev. Lett. \textbf{76}, 2782 (1996).

\bibitem{halperin_83}
B. I. Halperin, Helv. Phys. Acta \textbf{56}, 75 (1983).

\bibitem{qiu_89}
X. Qiu, R. Joynt, and A. H. MacDonald, Phys. Rev. B \textbf{40}, 11943 (1989).

\bibitem{jaud_00}
J. D. Naud, L P. Pryadko, and S. L. Sondhi, Phys. Rev. Lett. \textbf{85}, 5408 (2000).

\bibitem{levin_09}
M. Levin and M. P. A. Fisher, Phys. Rev. B \textbf{79}, 235315 (2009).

\bibitem{wang_13}
C. Wang and T. Senthil, Phys. Rev. B \textbf{87}, 235122 (2013).

\bibitem{jian_14}
C.-M. Jian and X.-L. Qi, Phys. Rev. X \textbf{4}, 041043 (2014).

\bibitem{wang_14}
C. Wang and M. Levin, Phys. Rev. Lett. \textbf{113}, 080403 (2014).

\bibitem{wang_15}
J. C. Wang and X.-G. Wen, Phys. Rev. B \textbf{91}, 035134 (2015).

\bibitem{Jiang_14}
S. Jiang, A. Mesaros, and Y. Ran, Phys. Rev. X \textbf{4}, 031048 (2014).

\bibitem{moradi_15}
H. Moradi and X.-G. Wen, Phys. Rev. B \textbf{91}, 075114 (2015).

\bibitem{sondhi_00}
S. L. Sondhi and K. Yang, Phys. Rev. B \textbf{63}, 054430 (2001).

\bibitem{kane_02}
C. L. Kane, R. Mukhopadhyay, and T. C. Lubensky, Phys. Rev. Lett. \textbf{88}, 036401 (2002).

\bibitem{teo_14}
J. C. Y. Teo and C. L. Kane, Phys. Rev. B \textbf{89}, 085101 (2014).

\bibitem{lu-12prb125119}
Y.-M. Lu and A. Vishwanath, Phys. Rev. B {\bf 86},  125119  (2012).

\bibitem{vaezi-14prl236804}
A. Vaezi and M. Barkeshli, Phys. Rev. Lett. {\bf 113},  236804  (2014).

\bibitem{seroussi-14prb104523}
I. Seroussi, E. Berg, and Y. Oreg, Phys. Rev. B {\bf 89},  104523  (2014).

\bibitem{klinovaja-14epjb87}
J. Klinovaja and D. Loss, Eur. Phys. J. B {\bf 87},  171  (2014).

\bibitem{meng-14epjb203}
T. Meng, P. Stano, J. Klinovaja, and D. Loss, Eur. Phys. J. B {\bf 87},  203 (2014).

\bibitem{sagi-14prb201102}
E. Sagi and Y. Oreg, Phys. Rev. B {\bf 90},  201102  (2014).

\bibitem{klino_14}
J. Klinovaja and Y. Tserkovnyak, Phys. Rev. B \textbf{90}, 115426 (2014).

\bibitem{mong-14prx011036}
R. S. K. Mong, D. J. Clarke, J. Alicea, N. H. Lindner, P. Fendley, C. Nayak, Y. Oreg, A. Stern, E. Berg, K. Shtengel, and M. P. A. Fisher, Phys. Rev. X {\bf 4},  011036  (2014).

\bibitem{vaezi14prx031009}
A. Vaezi, Phys. Rev. X {\bf 4},  031009  (2014).

\bibitem{klino_15}
J. Klinovaja, Y. Tserkovnyak, and D. Loss, Phys. Rev. B \textbf{91}, 085426 (2015).

\bibitem{santos_15}
R. A. Santos, C.-W. Huang, Y. Gefen, and D. B. Gutman, Phys. Rev. B \textbf{91}, 205141 (2015).

\bibitem{meng_15}
T. Meng, T. Neupert, M. Greiter, and R. Thomale, Phys. Rev. B \textbf{91}, 241106(R) (2015).

\bibitem{goroh_15}
G. Gorohovsky, R. G. Pereira, and E. Sela, Phys. Rev. B \textbf{91}, 245139 (2015).

\bibitem{neupert-14prb205101}
T. Neupert, C. Chamon, C. Mudry, and R. Thomale, Phys. Rev. B {\bf 90},  205101 (2014).

\bibitem{meng-14prb235425}
T. Meng and E. Sela, Phys. Rev. B {\bf 90},  235425  (2014).

\bibitem{burkov_balents_layers_11}
A. A. Burkov and L. Balents, Phys. Rev. Lett. \textbf{107}, 127205 (2011).

\bibitem{giamarchi_book}
T. Giamarchi, {\it Quantum Physics in One Dimension} (Oxford University Press, 2003).

\bibitem{charge_remark}
To determine the charge associated with a $2\pi$-kink in one of the bulk sine-Gordon terms, one can simply integrate the charge density given in Eq.~\eqref{eq:charge} over the kink, thus obtaining the quasiparticle charge.


\bibitem{meng_12}
T. Meng and L. Balents, Phys. Rev. B \textbf{86}, 054504 (2012).

\bibitem{volovik_book}
G. E. Volovik, \textit{The Universe in a Helium Droplet} (Clarendon Press, Oxford, 2003).


\bibitem{delf_schoeller_review}
J. von Delft and H. Schoeller, Annalen Phys. \textbf{7} 225, 1998.

\bibitem{sagi_preprint}
E. Sagi and Y. Oreg, arXiv:1506.02033.

\bibitem{wang_15_fract_ws}
Z. Wang, Physica B \textbf{475}, 80 (2015).


\bibitem{soi_coldatoms_1}
P. Wang, Z.-Q. Yu, Z. Fu, J. Miao, L. Huang, S. Chai, H. Zhai, and J. Zhang Phys. Rev. Lett. \textbf{109}, 095301 (2012).

\bibitem{soi_coldatoms_2}
L. W. Cheuk, A. T. Sommer, Z. Hadzibabic, T. Yefsah, W. S. Bakr, and M. W. Zwierlein, Phys. Rev. Lett. \textbf{109}, 095302 (2012).

\bibitem{cold_atoms_lattices}
P. Windpassinger and K. Sengstock, Rep. Prog. Phys. \textbf{76}, 086401 (2013).

\bibitem{cold_atoms_unit_cell}
S. Nascimb\`{e}ne, Y.-A. Chen, M. Atala, M. Aidelsburger, S. Trotzky, B. Paredes, and I. Bloch, Phys. Rev. Lett. \textbf{108}, 205301 (2012).



\end{thebibliography}
\end{document}